\documentstyle[aps,prl,epsfig]{revtex}
\begin{document}

\title{Electron fractionalization induced dephasing in Luttinger liquids} 
\author{Karyn Le Hur} 
\address{D\'epartement de Physique Th\'eorique, Universit\'e de Gen\`eve,
CH-1211, Gen\`eve 4, Switzerland.} 
\maketitle 

\begin{abstract} 
Using the appropriate fractionalization mechanism, we correctly derive
the temperature (T) and interaction dependence of the electron lifetime
$\tau_F$ in Luttinger liquids. For strong enough interactions, we report that 
$(T\tau_F)\propto g$, with $g\ll 1$ being
the standard Luttinger exponent; This reinforces that 
electrons are {\it not} good quasiparticles.
We immediately emphasize that this 
is of importance for the detection of electronic
interferences in ballistic 1D rings and carbon nanotubes, 
inducing ``dephasing'' (strong reduction of Aharonov-Bohm oscillations).
\end{abstract}

\vskip 0.5cm


(anyons, composite fermions, Luttinger liquid, etc.)

\vskip 0.5cm

{\bf Introduction.---} 
Recently, Pham {\it et al.} reconsidered the spectrum of the 
Luttinger liquid (LL), originally understood in terms of plasmons
(density fluctuations)\cite{L}, and within the context of 
bosonization gave an alternative and precise
description in terms of ``fractional quasiparticles''\cite{K.V}. 
These correspond to genuine excitations of LLs 
with non-zero {\it charge} or/and {\it current} with respect to the ground 
state (with no plasmon excited). The most famous
example is the charge-1/3 excitation
for the fractional Quantum Hall effect
at filling $\nu=1/3$, 
revealed in resonant tunneling 
and shot-noise type experiments\cite{Samin}. These could also be 
observed in carbon nanotubes\cite{Bena}. 
The current of one-dimensional (1D) quantum wires is naturally
carried by fractional modes\cite{current,Safi}. More precisely, if one
injects an {\it extra} spinless
right-moving electron at a time $t=0$
at a point $x=0$, this will irrefutably decompose
into two counter-propagating modes (the true quasiparticles), namely
a charge $Q_+=(1+g)/2$ state going to the right at the {\it plasmon} velocity 
$u$ and a charge $Q_-=(1-g)/2$ state going to the left at the 
velocity $-u$;
$g<1$ is the Luttinger parameter depending on interactions between
electrons\cite{K.V,Safi}. An electron does not propagate at the Fermi velocity.

Here, we are concerned by deriving 
an appropriate definition of the electron lifetime $\tau_F$
in LLs (at a finite temperature T), 
taking into account that ``fractionalization'' 
is the main source of (intrinsic) electron decoherence\cite{Apel}. 
We already stress that $\tau_F$ is the time needed for
the propagator of the counter-going mode $Q_-$ to vanish at the
{\it position} of the right-going mode $Q_+$, i.e., at 
$x\approx u\tau_F$: The probability
amplitude to find the electron -i.e., the two modes simultaneously- at
a distance close to the decoherence length $L_F=u\tau_F$, then
{\it vanishes}. We will show that
\begin{equation}
\tau_F^{-1}=\pi T\hbox{\Large{(}}\frac{g+g^{-1}}{2}-1\hbox{\Large{)}}.
\end{equation}
For strong enough interactions, we find
$(T\tau_F)\ll 1$ yielding that an
electron is not a good (Landau) quasiparticle in 1D.
Remember the main difference with a two-dimensional (2D) Fermi liquid, 
where $\tau_F$
is rather the time over which a thermally
excited electron (here, a Landau quasiparticle) simply
loses its phase coherence due to {\it collisions} with other 
electrons; In 2D, $\tau_{F}^{-1}\propto T^2\ln T$ nicely statisfies the
standard Fermi liquid criterion
$(T\tau_F)\gg 1$. The electron lifetime will be also investigated in the case
of fermions with spin.

This work is relevant for the
understanding of Aharonov-Bohm (AB) type experiments in mesoscopic
1D interferometers.  
We show that the probability of electronic interferences through
a ballistic ring with a four-terminal geometry, 
described by a Luttinger theory, will be reduced by a factor 
$\exp(-2\hat{\tau}/\tau_{\phi})$ where
$\tau_{\phi}=\tau_F$; $\hat{\tau}$
is the time that an electron spends in each arm 
of the ring. Like for 2D (clean) Fermi liquids\cite{Yac},  
the ``dephasing'' time $\tau_{\phi}$ 
is a key parameter determining the
electron lifetime; {\it We keep the traditional word dephasing time
to name ${\mathit \tau_{\phi}}$ in the formula above even though in LLs the
latter is related to the electron fractionalization 
(orthogonality catastrophe) and not to loss of phase coherence}\cite{AAK}. 
Recent experiments on 1D small 
ballistic rings\cite{Ha} and on multi-walled
carbon nanotubes in the quasi-ballistic regime\cite{Schon} 
report a phase-breaking time
$\tau_{\phi}\propto 1/T$, in agreement with our prediction.

{\bf Fractional Quasiparticles.---} 
First, we consider spinless fermions in 1D (e.g., electrons in strong 
magnetic field).
The whole physics is embodied in the Luttinger bosonized Hamiltonian\cite{L}
\begin{equation}
H_{Lut}=\frac{u}{2}\int_0^L dx\ \frac{1}{g}(\partial_x\phi)^2+
g(\partial_x\theta)^2.
\end{equation}
The electron system of size $L$ is first
governed by density fluctuations ({\it bosons}), or plasmons.
The mode $\phi$ is related to the electron density
$(\rho-\rho_o)=-\partial_x\phi/\sqrt{\pi}$ ($\rho_o$ is the mean density) 
and $\theta$ is the usual superfluid phase. 
The parameter $u=v_F/g$ depicts the plasmon velocity
and $g$ measures the strength of the interaction
between electrons; For free electrons $g=1$ $(u=v_F)$ and for repulsive
interactions $g<1$. But, in fact, we have 
the more generic expansions\cite{L}:
\begin{eqnarray}
\theta(x) &=& {\theta}_0+\frac{\sqrt{\pi}}{L}Jx+
\frac{1}{\sqrt{L}}\sum_{q\neq 0}
\theta_q \exp i qx,\\ \nonumber
\phi(x) &=&{\phi}_0-\frac{\sqrt{\pi}}{L}Qx + 
\frac{1}{\sqrt{L}}\sum_{q\neq 0}
\phi_q \exp i qx,
\end{eqnarray}
where $J$ and $Q$ (integral) describe the current and the charge
which are eventually injected in the (electron) system above the 
ground state: This also allows fractional 
{\it non-bosonic} excitations in LLs which can be nicely identified
using the chiral separation $\Theta_{\pm}=\theta\mp\phi/g$\cite{K.V}; 
The index
$\pm$ refers to the {\it direction of propagation} (right/left).
Then,
\vskip -0.1cm
\begin{equation}
H_{Lut}=H_+ +H_- = \frac{ug}{4}\int_0^L dx\ (\nabla\Theta_+)^2+(\nabla\Theta_-)^2,
\end{equation}
with $[H_+,H_-]=0$. The equations of motion $u\partial_x
\Theta_{\pm}=\mp\partial_t\Theta_{\pm}$ guarantee  
$\Theta_{\pm}(x,t)=\Theta_{\pm}(x\mp ut)$.
From Eq.(4), the non-bosonic
part of the Hamiltonian reads
$H_{\pm}=(u\pi/Lg){Q_{\pm}}^2$ with $Q_{\pm}=(Q\pm gJ)/2$
and the {\it true quasiparticles} -associated to $Q\neq 0$ and/or $J\neq 0$- 
are simply described by the (chiral vertex) operators
\begin{equation}
\label{fraco}
{\cal L}_{\pm}^{Q_{\pm}}(x,t)=\exp [-i\sqrt{\pi}Q_{\pm}\Theta_{\pm}(x,t)].
\end{equation}
Their Fourier transforms are indeed exact eigenstates of $H_{\pm}$\cite{K.V}. 
Remember that in LLs, a quasiparticle 
carries an {\it irrational} charge $Q_{\pm}=(Q\pm gJ)/2$. 
In pure current
processes (i.e., $Q=0$) we get $Q_{\pm}=\pm gJ/2$. 
This traduces that
$J/2$ pairs of Laughlin quasiparticle and quasihole with charge $\pm g$ 
and velocity $u$, are generated and the true current reads 
$J_R=g(uJ)$\cite{current}. 
Such quasiparticles with charge-1/3 have 
been revealed using resonant tunneling and shot-noise experiments in 
Quantum Hall fluids for 
filling factor $\nu=g=1/3$\cite{Samin}. 
Below, we are rather
interested in other fractionally-charged excitations.
We add an extra right-going spinless electron 
at the position $x=0$ at the time $t=0$: This
is a mixed $Q=1$ (charge) and $J=1$ (current) 
excitation, with no plasmon excited. 
According to Eq.(\ref{fraco}), this electron decomposes into
a charge $Q_+=(1+g)/2$ state going to the
right and a charge $Q_-=(1-g)/2$ state moving to the  
left. This ``splitting'' process guarantees a short 
decoherence time for the 
electron at {\it finite} temperature. 

{\bf Electron lifetime.---} Now, we precisely build the 
latter in an appropriate manner.
First, the electron operator at $x=0$ at $t=0$ must be rewritten in terms 
of the {\bf good} quasiparticles
\begin{equation}
\Psi_+^{\dagger}(0,0)=\exp \hbox{\Large{[}}-i\sqrt{\pi}(-\phi+\theta)(0,0)
\hbox{\Large{]}}=
{\cal L}_+^{\frac{1+g}{2}}(0,0)\cdot{\cal L}_-^{\frac{1-g}{2}}(0,0).
\end{equation}
We check that the electron is a good
eigenstate {\it only} for $g=1$: $Q_+=1$ and $Q_-=0$.
Below, we investigate how the wave packets
associated to the two counter-propagating ``fractons'' evolve in
space and time. For that issue, we compute their Green's functions.
At a finite temperature T $(\beta=1/T)$, we find the following
bosonic correlators
\vskip -0.1cm
\begin{equation}
<\Theta_{\pm}(x,\tau)\Theta_{\pm}(0,0)-\Theta_{\pm}(x,\tau)^2>\ 
=\frac{1}{g\pi}\ln\hbox{\Huge{(}}
\frac{\pi/(u\beta)}{\sin(
\frac{\pi}{\beta}[\tau\mp i\frac{x}{u}])}\hbox{\Huge{)}},
\end{equation}
\vskip -0.1cm
where $\tau=it$ is the usual Matsubara time
(We neglect finite size effects). This results in
\begin{eqnarray}
<({\cal L}_+^{\frac{1+g}{2}})^{\dagger}(x,\tau)
{\cal L}_+^{\frac{1+g}{2}}(0,0)>\hskip -0.1cm &=& \exp\hbox{\huge{(}}
\pi\frac{(1+g)^2}{4}<\Theta_+(x,\tau)\Theta_+(0,0)-
\Theta_+(x,\tau)^2>\hskip -0.1cm\hbox{\huge{)}}\\ \nonumber
&=& \hbox{\Large{[}}\frac{\pi}{(u\beta)}\hbox{\Large{]}}^{\gamma+1}
\hbox{\Large{[}}\sin(\frac{\pi}{\beta}[\tau-i\frac{x}{u}])\hbox{\Large{]}}
^{-\gamma-1}.
\end{eqnarray}
For convenience, we have redefined $\gamma=-1/2+(g+g^{-1})/4>0$. 
Similarly, we get
\begin{equation}
\label{count}
<({\cal L}_-^{\frac{1-g}{2}})^{\dagger}(x,\tau)
{\cal L}_-^{\frac{1-g}{2}}(0,0)>\ = 
\hbox{\Large{[}}\frac{\pi}{(u\beta)}\hbox{\Large{]}}^{\gamma}
 \hbox{\Large{[}}\sin(\frac{\pi}{\beta}[\tau+i\frac{x}{u}])\hbox{\Large{]}}
^{-\gamma}.
\end{equation}
This provides us a clear understanding of the right-going electron
Green's function\cite{L}, as:
\begin{equation}
<\Psi_+(x,t)\Psi_+^{\dagger}(0,0)>\ =\
<({\cal L}_+^{\frac{1+g}{2}})^{\dagger}(x,t){\cal L}_+^{\frac{1+g}{2}}(0,0)>
\times<
({\cal L}_-^{\frac{1-g}{2}})^{\dagger}(x,t){\cal L}_-^{\frac{1-g}{2}}(0,0)>.
\end{equation}
This identification is in fact 
essential to correctly evaluate the electron lifetime in LLs; See note in 
\cite{note-s}. 

More precisely, the right-moving fractional charge $Q_+=(1+g)/2$ 
propagates at the plasmon sound
velocity $u$. Its Green's function indeed possesses a resonance
for $\tau=ix/u$ $(t=x/u)$. Thus, in a time $t>0$,
$Q_+$ will be located at a certain position $x\approx ut$. On the other hand, 
the probability amplitude that the
counter-propagating charge $Q_-=
(1-g)/2$ reaches the {\bf same} position
$x>0$ (close to $ut$) at the time $t$, decreases 
exponentially with time at {\it finite} temperature:
\begin{equation}
\label{frac}
<({\cal L}_-^{\frac{1-g}{2}})^{\dagger}(x,t)
{\cal L}_-^{\frac{1-g}{2}}(0,0)>|_{x\rightarrow ut>0}
\propto \hbox{\Large{[}}
\sinh(\frac{\pi}{\beta}2t)\hbox{\Large{]}}^{-\gamma}
\approx\exp(-\frac{\pi}{\beta}2\gamma t).
\end{equation}
Such a decay for $x\approx ut>0$ 
simply traduces that the mode $Q_-$ goes to the left.
Using Eq.(10), we find that at the time $\tau_F$, defined as 
\begin{equation}
\label{lt}
\tau_F^{-1}=\frac{\pi}{\beta}2\gamma=-\partial_t \ln 
<({\cal L}_-^{\frac{1-g}{2}})^{\dagger}(x,t)
{\cal L}_-^{\frac{1-g}{2}}(0,0)>|_{x\rightarrow ut},
\end{equation}
the probability amplitude to recombine the  
electron at a point close to the 
decoherence $L_F=u\tau_F$, vanishes\cite{note1}:
\begin{equation}
\label{green}
G_+\hbox{\large{(}}x\rightarrow ut,t\hbox{\large{)}}\ 
=\ <\Psi_+(x,t)\Psi_+^{\dagger}(0,0)>|_{x\rightarrow ut}\ \approx \ 
e^{-t/\tau_F}\times
G_+^o\hbox{\large{(}}x\rightarrow ut,t\hbox{\large{)}}.
\end{equation}
Here, $G_+^o$ refers to the Green's function in the 
absence of fractionalization. {\it 
Remember that the temporal (exponential) decay
of the overlap between the right moving and the left moving fractons is taken
to define the electron lifetime} ${\mathit \tau_F}$.
In the above equation, we get the constraint 
$x\rightarrow ut$ highlighting the fact that 
fractional charges propagate at the plasmon 
velocity, and not at the Fermi velocity $v_F$\cite{Apel}.
This is equivalent to
 rewrite $\Psi_+(x,t)=\sum_{q} c_q(t)e^{-t/\tau_F}e^{iqx}$; 
the annihilation operator $c_q(t)$ 
evolves like in a Fermi gas\cite{note1} and 
the sum is restricted to the states in the vicinity of the 
Fermi level. Each 
level-q close to the Fermi level then acquires the same decoherence
time: $\tau_F=(2\pi\gamma T)^{-1}$. This is the correct
lifetime for a spinless electron in LLs.    

{\bf Application.---} Now, we like to repeat that this work is of special
importance for the analysis of AB type experiments in ballistic 1D
interferometers. For simplicity, a four-terminal geometry is assumed and 
contacts are supposed to be perfect\cite{M}.
Here, an injected electron 
can enter {\it either} of the two arms of the ring, and then may 
interfere with itself; 
To compute the probability of \underline{electronic} interferences
through such 
a ballistic ring we can simply use the following Landauer type formula
rewritten in our many-body language: ${\cal T}(\Phi)=2\Re e
\hbox{\large{(}}G_{+1}^*(L,\hat{\tau})G_{+2}(L,\hat{\tau})
\exp(2i\pi\Phi/\Phi_o)\hbox{\large{)}}$, 
where $\Phi$ is the 
magnetic flux enclosed to the ring and $\Phi_o$ is the flux quantum.
The propagator $G_{+j}(L,\hat{\tau})$ naturally corresponds to the
{\it electron transmission amplitude} (at the Fermi energy) 
through each arm $(j=1,2)$ of size L in the
{\it semi-classical} time $\hat{\tau}\approx L/u$. We neglect dynamical 
effects assuming $T\gg\omega$ (frequency). We have absorbed the
magnetic flux in a standard manner by rescaling
$G_{+j=1,2}(L,\hat{\tau})\rightarrow \exp(\mp i\pi\Phi/\Phi_o)G_{+j}
(L,\hat{\tau})$. This takes into account the 
phase accumulated by the total electron in 
the ring\cite{L}. When each arm of the ring is described by
a {\it Luttinger theory} (due to interactions), 
then from Eq.(\ref{green}) this leads to $G_{+j}(L,\hat{\tau})\propto 
\exp(-\hat{\tau}/\tau_{\phi})=\exp(-L/L_{\phi})$
where $\tau_{\phi}=\tau_F$ and $L_{\phi}=L_F\propto 1/T$. Thus, due to 
electron fractionalization (in each arm),
${\cal T}(\Phi)$ will be diminished by a factor  
$\exp(-2\hat{\tau}/\tau_{F})=\exp(-2L/L_{F})$. The related conductance then
reads
\begin{equation}
{\cal G}(\Phi)=(e^2/h){\cal T}(\Phi)\propto e^{-2\pi\gamma T L/u}\cos
(2\pi\Phi/\Phi_o).
\end{equation}

\begin{figure}[ht]
\centerline{\epsfig{file=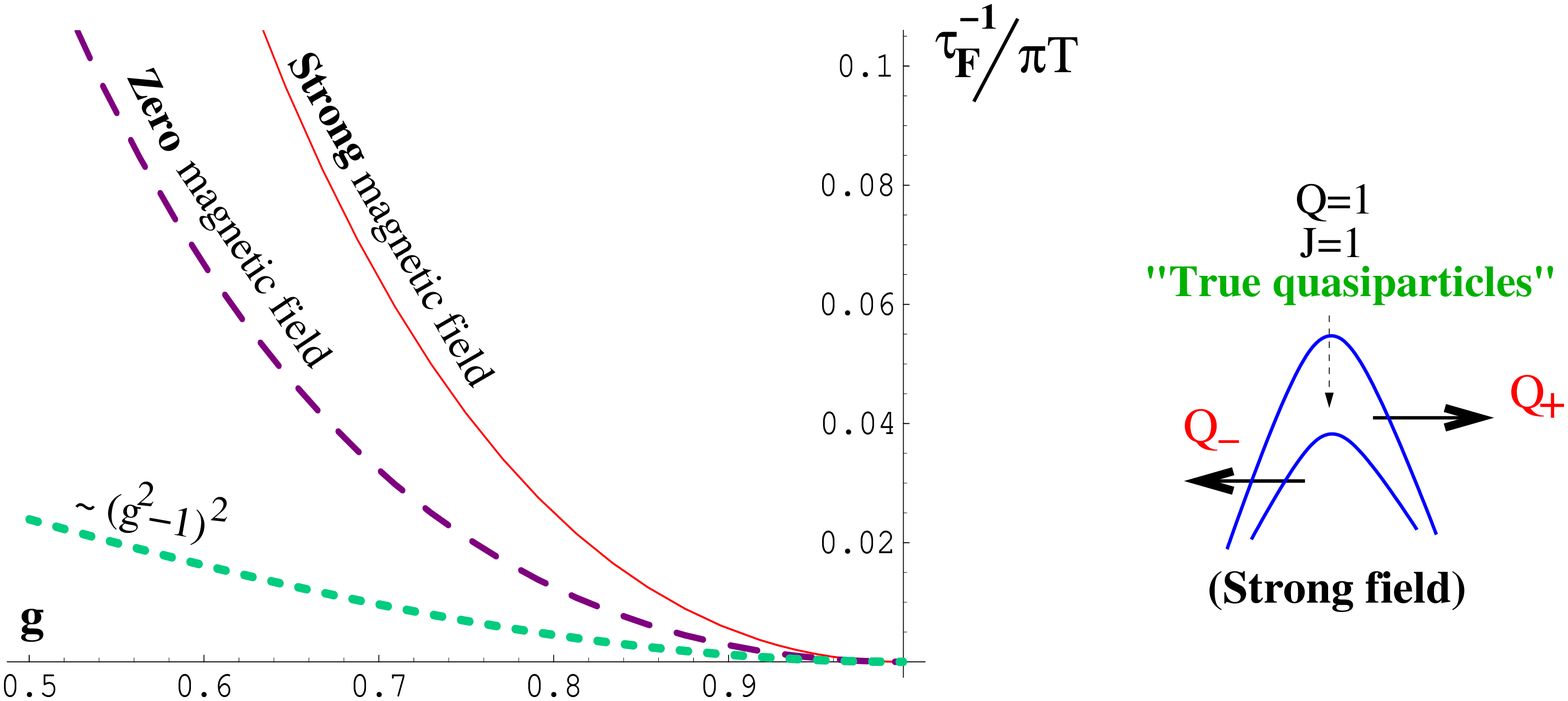,angle=0.0,height=5.5cm,width=10.1cm}}
\vskip 0.2cm
\caption{Inverse of the electron lifetime (normalized to $\pi T$) as
a function of the interaction parameter $g$, for zero- and strong magnetic
fields. We also draw the weak-coupling
approximation $(g\rightarrow 1)$ 
of the dephasing time recently computed in Ref.[14]
for a ballistic ring capacitively coupled to side-gates; Here, the Luttinger
exponent obeys $g^{-2}=1+e^2\nu_F/c$, where $c$ is the gate-arm 
capacitance per unit length and $\nu_F=1/(\pi v_F)$ the density of states
at the Fermi level [15].}
\end{figure}
\hskip -0.3cm 
Like for 2D (clean) Fermi liquids\cite{Yac}, the electron lifetime strongly 
reduces AB oscillations.
This may explain the recent experiments by 
Hansen {\it et al.} in ballistic 1D rings\cite{Ha} and 
by Sch\"{o}nenberger {\it et al.} in multi-walled carbon nanotubes in the 
quasi-ballistic regime\cite{Schon}. We agree on the 
approximation of the dephasing time made in Ref.\cite{M} 
for {\it weak} interactions between electrons (Fig.1). At $T=0$ the overlap 
between the two fractons which constitute the
electron only yields a power-law decay with time; See Eq.(\ref{frac}).
But, we cannot rigorously give a ruling on the behavior of ${\cal G}(\Phi)$
as a function of L since here we both neglect dynamical and finite
size effects.

{\bf Spinful case.---}
Let us now briefly discuss the situation of fermions
with spin. Here, the electron spectrum 
exhibits both spin-charge and chiral decomposition. The plasmon 
Hamiltonians exhibit the same form as in Eq.(2). We denote $u_s$ and $u_c$
the plasmon velocities for each sector (in general $u_c\neq u_s$)
while $g_c$ and $g_s$ are the two Luttinger parameters.
Fractional modes from the charge sector are now
 described by the operators
\begin{equation}
{C}_{\pm}^{Q_{c,\pm}}(x,t)=e^{-i{\sqrt\frac{\pi}{2}}
Q_{c,\pm}\Theta_{\pm}^c(x,t)},
\end{equation}
where $\Theta_{\pm}^c = \theta_c\mp {\phi}_c/g_c$. 
Spin excitations ${\cal S}_{\pm}^{Q_{s,\pm}}$ are defined in an identical
manner, and $Q_{c/s,\pm} = (Q_{\uparrow}+\tau Q_{\downarrow})/2 \pm g_c(
J_{\uparrow}+\tau J_{\downarrow})/2$; $\tau=+1$ for $Q_c$ and
$\tau=-1$ for $Q_s$\cite{K.V}. Suppose we add in a spin-up electron 
going to the right. This is
a mixed $Q_{\uparrow}=J_{\uparrow}=1$ and $Q_{\downarrow}=J_{\downarrow}=0$
excitation, which will then split into {\it four}
fractional parts. First, a right-moving 
charge $Q_{c,+}=(1+g_c)/2$ 
propagating at
velocity $+u_c$ and a right-moving spin (with a spin component) $S_+^z
=Q_{s,+}/2=(1+g_s)/4$ 
propagating at
velocity $u_s$. Second, a counter-propagating charge $Q_{c,-}=(1-g_c)/2$ 
with velocity $-u_c$ and a counter-propagating spin $S_-^z=(1-g_s)/4$
with velocity $-u_s$\cite{holon}. We can naturally identify
\begin{equation}
\Psi_{+\uparrow}^{\dagger}(0,0) 
=\hbox{\large{(}}{C}_+^{\frac{1+g_c}{2}}{\cal S}_+^{\frac{1+g_s}{2}}
{C}_-^{\frac{1-g_c}{2}}{\cal S}_-^{\frac{1-g_s}{2}}\hbox{\large{)}}(0,0).
\end{equation}
The lifetime for this spin-up electron going to the right is now 
defined as
\begin{equation}
e^{-t/\tau_F}\ =\ <({C}_-^{\frac{1-g_c}{2}})^{\dagger}(x,t)
{C}_-^{\frac{1-g_c}{2}}(0,0)>\times
<({\cal S}_-^{\frac{1-g_s}{2}})^{\dagger}(x,t)
{\cal S}_-^{\frac{1-g_s}{2}}(0,0)>|_{x\rightarrow u_ct \sim u_s t>0}.
\end{equation} 
We neglect corrections in ${\cal O}(u_c-u_s)$ assuming not too 
strong interactions. The probability amplitude to recombine the four 
``fractons'' (=the injected spin-up electron) at the {\bf same} point $x$ 
close 
to $u_c t\approx u_s t$, indeed vanishes for $t=\tau_F$.
A straightforward calculation gives
$\tau_F^{-1}\simeq\pi(\gamma_c+\gamma_s)/\beta$ where $\gamma_{\tau}=
(g_{\tau}+{g_{\tau}}^{-1})/4-1/2$. The decoherence
length $L_F$ is $L_F^{-1}\simeq\pi T(\gamma_c/u_c+\gamma_s/u_s)$. 
Equating $g_c=g_s=g$ and $u_c=u_s=u$, this corresponds to
situations in {\bf strong} magnetic field\cite{L}; We
 nicely reproduce the preceding result for spinless fermions.
At {\bf zero} magnetic field, one gets $g_s=1$ and $g_c=g$, and
$\tau_F$ becomes {\it maximum} because the spin part 
doesnot split due to SU(2) invariance (Fig.1).

{\bf Summary.---} We have, in this short paper, 
derived the {\it first} appropriate definition of the
electron lifetime in LLs based on the correct fractionalization
picture. For strong enough interactions,  
we find $(T\tau_F)=g\ll 1$: This really traduces
that the electron is {\it not} a good Landau 
quasiparticle. Approaching the aboslute zero, the electron lifetime drastically
increases reflecting that the wave packets of
the fractionally-charged modes become very extended.
We still emphasize that $\tau_F\propto 1/T$ 
can be detected, from AB type experiments,
in any {\it ballistic} 1D system described by
a {\it Luttinger theory}. We think that this provides a clear
understanding of the ``dephasing''
time(s) recently observed in small 1D ballistic rings\cite{Ha} and in
multi-walled carbon nanotubes above 10-20K\cite{Schon}. 
Again, although we use the word ``dephasing'' time, here the reduction
of AB oscillations must be related to the electron fractionalization and not
to loss of (electron) phase coherence. 
We will ponder the effect of coupling to an external environment 
elsewhere\cite{dissipation}. It would also
be worth to focus on the lifetime of the fractional quasiparticles
in LLs.

We would like to thank
M. B\"{u}ttiker, A. Hansen, 
L. Ioffe, K.-V. Pham, C. Sch\"{o}nenberger, and G. Seelig for discussions. 
This work was supported by the Swiss National Science 
Foundation.


\begin{references}
\bibitem{current} The current is $J_R=u(Q_+-Q_-)=g(uJ)$;
$uJ$ would be the current carried by the electrons. Applying a
difference of potential V 
between the two extremities of the sample produces
$(uJ)=V {\cal G}_o$ where ${\cal G}_o=e^2/h$. Then,
$J_R=V G$ with ${\cal G}=ge^2/h$.
\bibitem{note-s} In particular, one cannot use the spectral function 
$A(q,\omega)$ to compute the electron lifetime: By Fourier transforming the 
electron 
Green function, one does not get a quasiparticle peak but instead a 
complicated continuum with singularities.
\bibitem{note1} Here, 
$\hat{\Psi}_+(x,t)=\sum_{q} c_q(t) e^{iq x}$ refers to the 
operator of a right-mover
in the absence of fractionalization (but $u\neq v_F$). 
For $x\rightarrow ut$, we get the finite quantity
\begin{equation} 
G_+^o(x,t)=\ <\hat{\Psi}_+(x,t)\hat{\Psi}_+^{\dagger}(0,0)>\ \propto 
 \hbox{\Large{[}}\sinh(\frac{\pi}{\beta}[t-\frac{x}{u}])\hbox{\Large{]}}
^{-1}= [t-\frac{x}{u}]^{-1}
\approx\
<({\cal L}_+^{\frac{1+g}{2}})^{\dagger}(x,t){\cal L}_+^{\frac{1+g}{2}}(0,0)>.
\end{equation}
\bibitem{holon} In general, these are not the ``holon'' and
``spinon'' described by $C_{\pm}(Q_{c,\pm}=1)$ and 
${\cal S}_{\pm}(S_{\pm}^z=1/2)$\cite{K.V} 
(which are pure charge or spin process:
$J_{\uparrow}=J_{\downarrow}=0$!).
\end{references}
\end{document}